\documentclass[12pt,a4paper]{article}
\usepackage[latin1]{inputenc}
\usepackage{amsmath}
\usepackage{amsfonts}
\usepackage{amssymb}
\usepackage[left=2cm,right=2cm,top=2cm,bottom=2cm]{geometry}
\usepackage{tabularx}
\usepackage{subfig}
\usepackage{wrapfig}
\usepackage{tabularx}
\usepackage{sidecap}
\usepackage{setspace}
\usepackage{graphicx}
\usepackage{grffile}
\usepackage{placeins}
\usepackage{float}
\RequirePackage[colorlinks,citecolor=blue,urlcolor=blue,linkcolor=blue]{hyperref}
\begin{document}
\title{A model of mesons in finite extra-dimension
}
\author{Jugal Lahkar$^{1,a}$, D. K. Choudhury$^{1,2,}$,S.Roy$^{3}$and N.S. Bordoloi$^{4}$ \\ $^1$Department of Physics, Gauhati University, Guwahati-781 014, India. \\$^2$Center of theoretical studies,Pandu College,Guwahati-781012,India.\\$^3$Karimganj College,Karimganj-788710,Assam,India.\\$^{4}$Department of Physics,Cotton University,Guwahati-781001,India.\\ $^a$email:neetju77@gmail.com \\ }
\date{}
\maketitle
\doublespacing
\begin{abstract}
Recently,problem of stability of H-atom has been reported in extra-finite dimension,and found out that it is stable in extra-finite dimension of size,$R\leq\frac{a_0}{4}$,where,$a_0$ is the Bohr radius.  Assuming that,the heavy flavoured mesons have also such stability controlled by the scale of coupling constant,we obtain corresponding QCD Bohr radius and it is found to be well within the present theoretical and experimental limit of higher dimension.  We then study its consequences in their masses using effective string inspired potential model in higher dimension pursued by us. Our analysis suggests that an extra-dimension of size  $0.007\times10^{-15}m\leq L\leq 0.13\times10^{-15}m $  can account for the masses of heavy flavoured mesons under study within their experimental uncertainty,without confinement effect.This range is well below the present theoretical and experimental limits on the size of extra-dimension. 
\end{abstract}

\section{Introduction}
\label{Intro.}
\label{A}
There has been considerable discussion on the possibility of extra spatial dimension since long ago.  The idea was born in $1920$'s when Kaluza and Klein[1] for the first time introduced one additional dimension.  Later different theories such as ADD[2],RS[3],UED[4]etc were introduced supporting the idea.  In recent years,equally interesting research is going on what happens to the ordinary H-atom,when there is more than 3 dimension. The problem of stability of H-atom in a compact space was a subject of research in[5]by Bures and Seigl.  They reported that H-atom is stable in extra-finite dimension of size,$R\leq\frac{a_0}{4}$,where,$a_0$ is the Bohr radius.  Assuming that,the heavy flavoured mesons have also such stability controlled by the scale of coupling constant,we obtain corresponding QCD Bohr radius,$R_{QCD}\leq \frac{{a_0}\vert_{QCD}}{4}$,where ${a_0}\vert_{QCD}=\frac{3}{16\mu\alpha_s}$,$\alpha_s$ is the strong coupling constant and $\mu$ is the reduced mass of mesons.\\

Recently, a string inspired potential model[6],[7],[8] of mesons has been reported in higher dimension of infinite extent ,by us.  In this study we re-frame it for finite extra-dimension and study the masses of heavy-flavoured mesons. We assume the stability of mesons within their QCD Bohr radii in extra-dimension,and we study the masses of heavy flavoured mesons within this model. The additional assumption is that the modification of Coulomb potential in finite extra-dimension in a plausible way.  Plausible relationship between confinement effect and modified coulomb potential in finite extra-dimension is also suggested.\\
In section.2 we outline the formalism,in section.3 we calculate the masses of Heavy Flavoured mesons in this model without confinement,while section.4 is devoted to the Results for masses of heavy flavoured mesons.  The last section.5 includes conclusion and discussion.

\section{Formalism:}
\label{B}
\subsection{The string inspired potential model and its limitations:}
Recently a String inspired potential model[6-8] is reported which assumes Luscher term[9] to be the higher dimensional correction to Cornell potential[ 6-8],[10],
\begin{equation}
\label{E1}
v(r)=\frac{-\gamma}{r}+br+c
\end{equation}
where b is standard confinement parameter and
\begin{equation}
\label{E2}
\gamma=\frac{\pi(D-2)}{24}
\end{equation}
At $D=3$,
\begin{equation}
\label{E3}
\gamma=\frac{4\alpha_s}{3}
\implies\alpha_s=0.0981
\end{equation}
This coupling constant does not correspond to its numerical value for $D=3$.Following MS-bar scheme using standard equation of coupling constant,
\begin{equation}
{\alpha_s(Q^2)}=\frac{4\pi}{(11-\frac{2n_f}{3})(ln\frac{Q^2}{{\Lambda^2}_{(QCD)}})}
\end{equation}
The, value given in equation${(3)}$ is much smaller than the corresponding value $\alpha_s=0.39$ at c-scale and $\alpha_s=0.22$ at b-scale.  Similarly,for$n_f=6$,$\Lambda_{QCD}=0.382GeV$,the value of $\alpha{(Q)^2}$ of equation$(4)$ is reached at $Q^2=1.2\times{10^{13}}GeV^2$,while for $n_f=16,\Lambda_{QCD}=0.382GeV$ we get $Q^2\simeq5.7\times10^{(20)}Gev^2$.  So,the higher dimensional extension of Coulomb term in [6-8] does not correspond to standard QCD at D=3 in present energy regime and hence ruled out.  Below,we therefore suggest an alternative physically plausible prescription for it.
\subsection{Improved potential model in finite extra-dimension:}
We consider the known 3 dimensions to be in the range 0 to $\infty$ and the extra dimension to be finite within the range 0 to L[3,4,5]. Thus,
\begin{equation}
r_D^2 =r_1^2+r_2^2+r_3^2+y^2
\end{equation}
\begin{equation}
 =r^2+y^2
\end{equation}
where $r^2=r_1^2+r_2^2+r_3^2$ , y is the size of finite extra dimension.For $r>>y$ we get,
\begin{equation}
r_D\simeq r+\frac{y^2}{2r}
\end{equation}
Now,we consider the Coulomb potential in d-dimension,
\begin{equation}
V(r_D)=-\frac{A_D}{r_D}
\end{equation}
where,$A_D=\frac{4\alpha_s}{3}$,at D=3
And,with finite extra-dimension we modified it to,
\begin{equation}
\frac{4\alpha_s}{3}\longrightarrow\frac{4\alpha_s}{3}e^{{-\mu_{L}}{y}}
\end{equation}
Now,for ${\mu}_{L}y\ll1$,
\begin{equation}
A_D={\frac{4\alpha_s}{3}}{(1-\mu_{L} y)}
\end{equation}

At $D=3,y=0$,and we get back the standard 3-dimensional QCD coupling constant. The choice of equation $(9)$ is to indicate the short range measure of extra-dimension $y$ such that gluonic effect in the extra-dimension is of very short
range,of the order of $\lambda_L\approx\frac{1}{\mu_L}$. We have taken $\mu_L=1GeV$ in our analysis. But it is important to discuss the significance of our results if we vary $\mu_L$ which is basically considered as inverse of the range of gluon field in extra-dimension,which can't exceed QCD Bohr radii in our model. Obviously if $\mu_L$ increases, the value of $'L'$ will decrease,but in our model $'L'$ and $\frac{1}{\mu_L}$ can't increase beyond QCD Bohr radius.  \\

\subsection{Wave function(D-Dimensional,with only Coulomb term in finite extra dimension):}
The D-dimensional Schrodinger equation is[6-8],[13],
\begin{equation}
{[\frac{d^2}{dr_D^2}+\frac{D-1}{r_D}\frac{d}{dr_D}-\frac{l(l+D-2)}{r_D^2}+
\frac{2\mu}{\hbar^2}{(E-V_0)}]}R(r_D)=0
\end{equation}
for l=0,taking$\hbar=1$,we get
\begin{equation}
\ddot{R}(r_D)+\frac{D-1}{r_D}\dot{R}(r_D)+2\mu(E+\frac{A_D}{r_D})R(r_D)=0
\end{equation}
Let,$R(r_D)=F(r_D)e^{-\mu A_D r_D}$[6-8],[14],
Now putting $R(r_D)$ in equation $(12)$ we get,
\begin{equation}
\ddot{F}(r_D)+{(\frac{D-1}{r_D}-2\mu A_D)}\dot{F}(r_D)+{(\mu ^2A_D^2-\frac{D-1}{r_D}}\mu A_D+2\mu E+\frac{2\mu{A_D}{r_D})}F(r_D)=0
\end{equation}
Now,we consider the series expansion of $F(r_D)$ as,
$F(r_D)=\sum_{n=0}^{\infty}{a_nr_D^n}f(r_D,D)$,
such that $f(r_D)=1$ at $D=3$.Let us consider,$f(r_D)=r^{\frac{\sigma{(D-3)}}{2}}$[15],which satisfies this condition.Then the radial wave function can be expressed as,
$R(r_D)=\sum_{n=0}^{\infty}{a_nr_D^{n+\frac{\sigma{(D-3)}}{2}}}e^{-\mu A_D r_D}$.
For ground state,$n=0$,we get the unperturbed wave function,
\begin{equation}
\psi(r_D)=N_D(r^2+y^2)^\frac{\sigma{(D-3)}}{2}e^{-\mu A_D (r+\frac{y^2}{2r})}
\end{equation}
Now,at $D=3,y=0$ and we get from above equation$(14)$,
\begin{equation}
\psi(r)=Ne^{-\mu \frac{4\alpha_s}{3} r}
\end{equation}
which is consistent with standard H-atom wave function[15] at D=3.
\subsection{Normalization: with 3 non-compact and one compact extra dimension}
The normalization condition[16] is,
\begin{equation}
\int_{L}^{\infty}\int
_{0}^{L}DC_D(r^2+y^2)^{\frac{(D-1)}{2}}\vert\psi(r,y)\vert^2 drdy=1
\end{equation}
where,$C_D=\frac{(\pi)^{\frac{D}{2}}}{\Gamma{(\frac{D}{2}+1)}}$.Now substituting the wave-function we get,for $D=4$,
\begin{equation}
DC_DN_D^2\int_{L}^{\infty}\int
_{0}^{L}r_D^{\sigma +3}e^{-2\mu A_D r_D} drdy=1
\end{equation}
Hence,it is clear from above equation that $\sigma$ and $N_D$ are interrelated. In 3-dimension $\sigma$ do not have any physical significance.  For any given value of $'\sigma'$ one can find $'N_D'$ at $D=4,5,6,...........$  etc. After some calculation,neglecting higher order terms of $'L'$,we get,
\begin{equation}
N_D=[\frac{K^5}{DC_DL{\Gamma(5)}}]^{\frac{1}{2}}
 \end{equation}
 Where,$K=2\mu A_D$ and $A_D=\frac{4\alpha_s}{3}(1-\mu_Ly)$
 \begin{table}[h]
\caption{Normalization constant in $D=4$ for $\sigma=1$:}
\label{table1}
\begin{center}
\begin{tabularx}{10cm}{|X|X|}
\hline
Mesons & $N_D$  \\ \hline
$D{(c\overline{s})}$ & $0.01$     \\ \hline
$D{(c\overline{d})}$ & $0.005$     \\ \hline
$B{(u\overline{b})}$ & $0.0015$     \\ \hline
$B{(s\overline{b})}$ & $0.003$       \\ \hline
$B{(\overline{b}c)}$ & $0.12$         \\ \hline

\end{tabularx}
\end{center}
\end{table} 
\pagebreak 
\section{Masses of heavy flavoured mesons in compact extra-dimension without confinement:}
\subsubsection{Mass only with coulomb term in compact extra dimension:}
As a application of the formalism developed in $2$, we calculate the masses of Heavy flavoured mesons.Pseudo-scalar meson mass can be computed from the following relation[12],[17]:
\begin{equation}
M_p=m_Q+m_{\overline{Q}}+\Delta E
\end{equation}
where,$\Delta E=\langle H \rangle $.
In D-spatial dimension,the Hamiltonian operator H has the form[6],[18],[19]:
\begin{equation}
H=-\frac{\nabla_D^2}{2\mu}+V(r_D)
\end{equation}
where,$\mu = \dfrac{m_Qm_{\overline{Q}}}{m_Q+m_{\overline{Q}}}$ is the reduced mass of the meson with $m_Q$ and $m_{\overline{Q}}$ are the quark and anti-quark masses;$V(r_D)$ corresponds to the coulomb plus confinement term with plausible modification to compact extra dimension given as
\begin{equation}
V(r_D)=-\frac{A_D}{r_D}+br_D
\end{equation}
In this work we first neglect the confinement term and see the effect of coulomb term with modification in compact extra dimension,and $\nabla^2_D$ is the Laplace's operator in D- dimension[19], which at $l=0$ is given by,
\begin{equation}
\nabla^2_D\equiv \frac{d^2}{dr_D^2}+{\frac{D-1}{r_D}}{\frac{d}{dr_D}}
\end{equation}
Now,$\langle H \rangle$ can be expressed as(with only Coulomb term in the potential in compact extra dimension),
\begin{equation}
\langle H \rangle =\langle -\frac{\nabla_D^2}{2\mu} \rangle+\langle \frac{-A_D}{r_D} \rangle
\end{equation}

\begin{equation}
\langle H \rangle =\langle -\frac{\nabla^2_3}{2\mu} \rangle+\langle -{\frac{1}{2\mu}}{\frac{\delta^2}{\delta y^2}} \rangle+\langle -\frac{A_D}{r_D} \rangle
=\langle H_1 \rangle +\langle H_2 \rangle +\langle H_3 \rangle
\end{equation}
where first two terms are the kinematic contribution and third term corresponds to potential contribution.Now putting  D-dimensional wave function  given in(14) and using coupling constant from equation $(10)$ in equation $(24)$ we get,

\begin{equation}
\langle H_1 \rangle=\langle -\frac{\nabla^2_3}{2\mu} \rangle = \frac{9}{32\mu}{{(\frac{\mu\pi}{12})}^2}
\end{equation}
For the only compact extra dimension,$\psi=N_D e^{-\mu A_D y}$,and with it we get,
\begin{equation}
\langle H_2 \rangle=\langle -{\frac{1}{2\mu}}{\frac{\delta^2}{\delta y^2}} \rangle=\frac{N_D^2 A_D}{4}{(1+2\mu A_DL)}
\end{equation}
And the potential term,
\begin{equation}
\langle H_3 \rangle=\langle -\frac{A_D}{r_D} \rangle=N_D^2 A_D D C_D{[\frac{1}{2}\frac{\sqrt{\pi}}{\sqrt{\mu A_D}}\frac{\Gamma{(5\sigma+\frac{1}{2})}}{{(2\mu A_D)}^{(5\sigma+\frac{1}{2})}}+\frac{1}{2}\sqrt{\pi}L\frac{\Gamma{(5\sigma)}}{{(2\mu A_D)}^{5\sigma}}]}
\end{equation}
neglecting higher orders of L,since L is very small.Then we get the final result,

 \begin{equation}
 \langle H \rangle=\frac{9}{32\mu}{{(\frac{\mu\pi}{12})}^2}+\frac{N_D^2 A_D}{4}{(1+2\mu A_DL)}+N_D^2 A_D D C_D{[\frac{1}{2}\frac{\sqrt{\pi}}{\sqrt{\mu A_D}}\frac{\Gamma{(5\sigma+\frac{1}{2})}}{{(2\mu A_D)}^{(5\sigma+\frac{1}{2})}}+\frac{1}{2}\sqrt{\pi}L \frac{\Gamma{(5\sigma)}}{{(2\mu A_D)}^{5\sigma}}]}
\end{equation}
We can write, 
\begin{equation}
\langle H \rangle=F({\mu})+G(\mu{L})
\end{equation}
where,$F{(\mu)}$ is L independent with $F{(\mu)}=\frac{9}{32\mu}{{(\frac{\mu\pi}{12})}^2}$and  \\ $G{({\mu}{L})}=\frac{N_D^2 A_D}{4}{(1+2\mu A_DL)}+N_D^2 A_D D C_D{[\frac{1}{2}\frac{\sqrt{\pi}}{\sqrt{\mu A_D}}\frac{\Gamma{(5\sigma+\frac{1}{2})}}{{(2\mu A_D)}^{(5\sigma+\frac{1}{2})}}+\frac{1}{2}\sqrt{\pi}L \frac{\Gamma{(5\sigma)}}{{(2\mu A_D)}^{5\sigma}}]}$,is L dependent. Since $G{({\mu}{L})}$ is not explicitly a linear function of L,equation$(29)$ shows that variation of mass with size of extra-dimension is not linear.\\

\section{Results:}
\subsection{Estimation of QCD Bohr radii of heavy flavoured mesons and comparision with various theory and experimental limits:}
In table$(2)$,we show that QCD Bohr radii of various heavy flavoured quark anti-quark systems in $"Metre(m)"$ and then compare with the experimental and theoretical limit of extra-dimension in table$(3)$.  As the obtained QCD Bohr radii are always well within the corresponding limits of table$(3)$,it is therefore,tempting to assume that physical mesons too might be stable within such QCD Bohr radii and see its consequences in their masses.
\begin{table}[h]
\caption{QCD Bohr radius}
\label{table1}
\begin{center}
\begin{tabularx}{16cm}{|X|X|X|}

\hline
Mesons & Reduced mass(Gev)&${a_0}\vert_{QCD}=\frac{3}{16\mu\alpha_s}(m)$  \\ \hline
    $B^0{(d\overline{b})} $ & 0.1733 &  $24.5\times10^{-15}$          \\ \hline
 
$B^0_s{(s\overline{b})}$ & 0.23727 &  $17.95\times10^{-15}$              \\ \hline
 $B^+_c{(c\overline{b})}$ &1.02254  &  $4.15\times10^{-15}$              \\ \hline

$D^-{(\overline{c}d)}$ & 0.15826     &   $ 15.15\times10^{-15}$             \\  \hline
$D^+_s{(c\overline{s})}$ & 0.2099    &    $11.45\times10^{-15}$               \\  \hline

\end{tabularx}
\end{center}
\end{table}

\begin{table}[h]
\caption{ Different experimental and theoretical limit on the size of extra dimension            }
\label{table2}
\begin{center}
\begin{tabularx}{12cm}{|X|X|}
\hline
Experiment and Models & Limit on the size of extra-dimension (m) \\ \hline
Fermi-LAT$[21]$ & $8\times10^{-9}m$  (LED)     \\ \hline
LEP-I$[20]$ & $4.5\times10^{-14}m$            \\ \hline
ADD $[2]$& $\sim10^{-3}m$                         \\ \hline
Martin Bures$[6]$ &$ \leq\frac{a_0}{4}{(0.13225\times10^{-10})m}$  \\ \hline
ALEPH,DELPHI,OPAL$[20]$ &$\sim6\times10^{-18} m$                          \\  \hline
RS$[3]$ &$2\times10^{-9}m $                                      \\ \hline
\end{tabularx}
\end{center}
\end{table} 

\subsection{Variation of mass of mesons with size of extra-dimension :}
With the expression obtained for $\langle H \rangle$,we calculate the mass of  heavy flavoured meson[6],[12], $B{(\overline{b}c)}$.  We take the value of $'L'$ according to the condition $R_{QCD}\leq\frac{3}{16\mu\alpha_s}$ [5].  In table$(4)$,we show the variation of mass of $B{(\overline{b}c)}$ meson with the size of finite extra-dimension,which is well within the corresponding QCD Bohr radii, $4.15fm$ {(table$(2)$)}. This table$(4)$ shows that as size of extra-dimension increases,mass of meson also increases.  Further,for L=0,at d=3,the calculated mass is above the experimental value suggesting that effectively confinement effect reduces mass of a meson in $3D$.  We therefore raise the question if such reduction of mass due to the confinement effect can be obtained equivalently through the assumption that gluon effect can as well propagate to extra-dimension of size within the QCD Bohr radii. From table$(4)$,it is clear that for $0.007\times10^{-15}m\leq L\leq0.009\times10^{-15}m$ the calculated mass of $B{(\overline{b}c)}$ meson comes closer to the experimental value.  Similar analysis is done for $\tau{(\overline{b}b)}$ in table$(5)$,and we find that for $0.1\times10^{-15}m\leq L \leq 0.13\times10^{-15}m$ the calculated mass comes closer to the experimental value.  The input parameters for numerical calculations used are $m_b=4.66Gev$,$m_c=1.31Gev$ and $\alpha_s$ values $0.39$ and $0.22$ for c-scale and b-scale respectively. Our analysis suggests that an additional extra-dimension of size $7\times10^{-18}m \leq L \leq 13\times10^{-17} m$ is sufficient to account for the masses of mesons under study without confinement effect.\\\\
\begin{table}[h]
\caption{Variation of mass of $B{(\overline{b}c)}$ meson with size of extra-dimension without confinement:}
\label{Table4}
\begin{center}
\begin{tabularx}{12cm}{|X|X|X|}
\hline
$L{(m)}$ & $ M_P{(Gev)} $ & $Exp. mass{(Gev)}$  \\ \hline

 $0.005 \times10^{-15} $& 6.24 &$ 6.27\pm0.006$      \\ \hline
$0.006   \times10^{-15} $          &6.247&      \\ \hline
$0.007   \times10^{-15} $           &6.25  &    \\ \hline
$0.008 \times10^{-15} $               &6.26   &  \\ \hline
$0.009   \times10^{-15} $                 &6.27& \\ \hline
$0.01   \times10^{-15} $                 & 6.4343&  \\ \hline
$0.02    \times10^{-15} $               & 6.4474   & \\ \hline
$0.03    \times10^{-15} $                   & 6.4578 &\\ \hline
$0.04     \times10^{-15} $                   & 6.469  & \\ \hline
$0.05      \times10^{-15} $                     & 6.4821 & \\ \hline

\end{tabularx}
\end{center}
\end{table}
\begin{table}[h]
\caption{Mass of $\tau{(\overline{b}b)}$meson without confinement in finite extra-dimension:}
\label{Table6}
\begin{center}
\begin{tabularx}{12cm}{|X|X|X|}
\hline
$Size{L}{(m)}$ &$M_P{(Gev)}$&$Exp.Mass(GeV)  $        \\ \hline
$0.05\times10^{-15}$        &9.439 &  $9.46\pm0.0022$   \\ \hline
$0.06  \times10^{-15}$                     &9.447&   \\ \hline
$0.07    \times10^{-15}$                   &9.45& \\ \hline
$0.08    \times10^{-15}$                       &9.453&   \\ \hline
$0.09   \times10^{-15}$                        &9.458&\\ \hline
$0.10   \times10^{-15}$                          &9.460 &     \\ \hline
$0.11    \times10^{-15}$                     &9.465  & \\ \hline
$0.12    \times10^{-15}$                         &9.468& \\ \hline
$0.13    \times10^{-15}$                        &9.47  & \\ \hline
$0.14     \times10^{-15}$                           &9.473 &\\ \hline
\end{tabularx}
\end{center}
\end{table}

\pagebreak
\section{Conclusion and Discussion:}  
In this paper,we have calculated the QCD Bohr radii for qurk-antiquark system of heavy flavoured mesons in analogy with the recent result that the H-atom can be stable in finite extra dimension if the size of extra dimension does not exceed $\frac{a_0}{4}$[5],where,$a_0$ is the Bohr radius.  It is found that the QCD Bohr radii of various Heavy Flavoured mesons are well within the present theoretical and experimental limit of size of extra-dimension.  We then study its consequences in their masses using an improved version of the effective string inspired potential model[6-8].  Within the uncertainty of masses of known Heavy Flavoured mesons the allowed range of extra dimension is $L\leq0.009\times10^{-15} m$ for $B_c$ meson and $L\leq0.13 \times10^{-15} m $ for $\tau$ meson.Our analysis suggests that an additional extra-dimension of size $7\times10^{-18}m \leq L \leq 13\times10^{-17} m$ is sufficient to account for the masses of mesons under study without confinement effect.  This size of finite extra-dimension (L) is well below the present theoretical and experimental limit, but far above the Planck length($\simeq 1.5\times10^{-35}$ m). \\\\
    Let us discuss the limitation of the model,there are differences between the bound state problem of hydrogen atom in extra-dimension and that of quark-antiquark system,while the potential of H-atom is purely coulombic but that of quark-antiquark system also contains confinement effect.We have therefore study the problem using an modified form of coulomb potential(equation 8),which effectively takes into account the confinement effect.There are also many cases where it has been established that there is no stable H-atom in extra-dimension. The dynamical condition of the stability or unstability of the quark-antiquark system is however beyond the scope of present model,which is basically based on QCD-QED analogy of Coulomb potential.
         
\section*{Acknowledgement}
One of the authors(D.K.C)thanks Prof.Mike Tepper,Rudolf Peirels Center of Theoretical Physics,University of Oxford, Prof.D.K.Srivastava of VECC,Kolkata,and Prof.P.Poulose ,IIT Guwahati,India for useful discussions.We thank  Dr. Bhaskar Jyoti Hazarika,Centre of theoretical Physics,Pandu College,Assam,India for collaboration at the initial stage of the work. One of the authors (J.L) acknowledges CSIR,India for financial support by providing Fellowship during the research work.


\begin{thebibliography}{10}

\bibitem{kaluza}T.Kaluza,Sitzungsber.Preuss.Akad.Wiss.Berlin(Math.Phys.)\textbf{K1},966(1921);O.Klein,Z.Phys.

\textbf{37},895(1926)
\bibitem{ADD}N.Arkani Hamed,S.Dimopolous,G.Dvali,Phys.Lett.B\textbf{429}, 263(1998)
\bibitem{RS}L.Randall,R.Sundrum,PRL,vol.\textbf{83}, 17(1999)
\bibitem{Appelquist}T. Appelquist, H.-C. Cheng and B.A. Dobrescu, Phys. Rev. D 64, 035002 (2001).
\bibitem{Bures}M.Bures and P.Seigl,Annals of Phy.,vol.\textbf{354},pp.316-327(2015)
\bibitem{roy}S.Roy and D.K.Choudhury.,Canadian J of Phys.,vol.94,\textbf{0549},1282-1288,(2016)

\bibitem{ROY}S.Roy and D.K.Choudhury.,Phys.scr.,\textbf{87},065101(9pp)(2013)
\bibitem{Sroy}S.Roy,D.K.Choudhury.,Phys.Scr.,\textbf{86}, 045101(6pp)(2012)

\bibitem{Luscher}Luscher etal.Nucl.Phys.,\textbf{B180},1(1981)
\bibitem{Eichten}Eichten et al.,Phys.Rev.,\textbf{D21}203(1980)
\bibitem{neubert}M.Neubert,Phy.Rept.,\textbf{245}, 259(1994)
\bibitem{rai}A.K.Rai,B.Patel and P.C.Vinodkumar,Phys.Rev.C\textbf{78}, 055202(2008)

\bibitem{dong}S.H.Dong,Wave equation in higher dimension,Springer,\textbf{81pp},2011,ISBN-978-94-007-1916-3
\bibitem{Roy}S.Roy and D.K.Choudhury,Quantum chromodynamics and string inspired potential model,(LAP Lambart Academic publishing,ISBN:978-3-659-70881-7)(2015)
\bibitem{beiser}A.Beiser,Concepts of Modern Physics,234pp,[ISBN:978-0-07-015155-0](2006)
\bibitem{Maura}Maura Eduarda and David G. Henderson.,Experiencing Geometry:On plane and sphere,Prentice Hall PTR,(1996)
\bibitem{Halzen}Quarks and Leptons by Halzen and Martin,(ISBN-13: 978-0471887416),2013 
\bibitem{oikonomou}V.K.Oikonomou and K.Kleids,Int.j.of Mod.Phys.A,vol.26(2011),4633-4666
\bibitem{dong}S.H.Dong et al.,Found.Phys.lett.,\textbf{12}, 465(1999)
\bibitem{Lep}Limits on the size of extra dimension,Particle physics and cosmology:second tropical workshop,\textbf{cp540},AIP,I-56396-965-3/00
\bibitem{Florates}Florates and Leonteries,Phys.Rev.Lett.\textbf{B465},(1999),95
\bibitem{Petreczky}Petreczky, P. Eur. Phys. J. C.\textbf{43: 51},(2005) doi:10.1140/epjc/s2005-02261-6



\end{thebibliography}
\end{document}